# Application of Inventory Management Principles for Efficient Data Placement in Storage Networks

**R Arokia Paul Rajan[1] and F Sagayaraj Francis[2]**

**[1] Department of Computer Science, Pope John Paul II College of Education, Pondicherry, India**

**[2] Department of Computer Science & Engineering, Pondicherry Engineering College, Pondicherry, India**

**Abstract**

The principles and strategies found in material management are comparable and analogue with the data management. This paper concentrates on the conversion of product inventory management principles into data inventory management principles. Efforts were made to enumerate various impacting parameters that would be appropriate to consider if any data inventory model could be plotted. The inventory models parameters are carefully tailored to fit in to the data management paradigm so as to match various distributed storage architectural systems. The basic function of storage management is to employ the storage resources efficiently so as to yield the maximum performance. Even though the storage medium is cheaper in cost that doesn't mean that addition or upgrading of the storage capacity alone can solve the problem of storing ever growing nature of data especially in the inter networked data centric applications. In such cases it is necessary to find a most fitting storage policy to bring out a significant improvement in the data retrieval performance. This paper emphasizes on the need for deriving generic principles for data inventory control that includes data inventory parameters which will be the base for proposing mathematical data inventory models.

## 1. Introduction

Inventory Management is a continued process of overseeing and controlling of ordering, storage and use of components that a company will use in production of the items it will sell as well as the overseeing and controlling of quantities of finished products for sale. Keeping the inventory level too high will lead to idle capital reposition without utilization and too little will result with costly interruptions. The best inventory policy should optimize the ordering level and capital investment at opt time.

Visualizing and adapting a best inventory strategic policy is a challenging task for an organization. .

A good inventory strategy sophisticates the administration to take better inventory control decisions. An inventory control decides and manages about when to replenish the items and how much it should be replenished. A good inventory policy [4] answers the following questions:

1. How much to order?
2. When to order?

These two questions are the biggest challenge thrown towards the top level management because if it is not viably answered the organization may end up with unnecessary lockage of capital or costly interruption to the business. These two questions are relatively influenced by many costs and these costs are called as Economic parameters.

There is a wide scope of improving the storage management principles [8] if the same is adapted with product inventory principles which are already bench marked.

## 2. Comparison of Product vs. Data inventory

There are many similarities that exist between a consumable product or commodity or services with data in the computer industry. The table below distinguishes between a product inventory and a data inventory.

Table 1: Comparison of Product and Data Inventory

| Product inventory | Data inventory |
|---|---|
| Items and components are the consumables | Data are the consumables. |
| Demand is Deterministic or Probabilistic. | Mostly Probabilistic in nature. |
| Demand may follow some pattern. | Pattern finding is difficult. |
| Costs heads are well predictable. | Costs heads may vary in nature influenced by external parameters. |
| Replications are pointless. | Replication [1] will have a major impact. |
| Reorder levels are well predictable. | Replication of data can be equated to the Reorder level. |
| Depreciation is natural. | Data will never lose its value. |
| Usage of items by users is limited. | Concurrent usage is possible. |
| Performance tuning is limited to upper bound. | Performance tuning is enhanceable. |
| Insurance policies are well defined. | Data insurance is difficult and subject to external uncontrollable factors. |
| Quality measures are well established. | Quality measures are tuneable. |
| Product is consumable only when it completes product life cycle. | Data is consumable at any stage. |
| Interruptions will end up with stoppage of the further process chain. | Interruptions can be handled with appropriate alternatives. |
| Request, Allot and Use are the primitive operations. | Read, Write, Share, Insert and Delete are the primitive set of operations. |

## 3. Related works

There are lots of data placement algorithms which are deployed as strategies for data placement in different distributed storage networks. The networks include SAN, NAS, P2P, Cluster, Grid and including Cloud [12],[13],[14],[15]. But the authors found there is no precedence of works that compares and applies the principles of product inventory principles to fit into the data placement principles.

## 4. Need for data inventory principle

Product inventory [6] is to manage efficiently the idle resources to the customer or process requests. If we fit this definition to the data inventory, we can coin: Data Inventory Management is to make it available or allocate optimum data resources fittingly in the optimized data stores to the demands of the users. A good product inventory control makes the items always available to the requests or needs of the customers or for further processes. Similarly data inventory control principle should be independent enough to manage the availability of the data to the users' requests with a good trade off of keeping the data at the best possible way to access with various influencing parameters.

### 4.1 Draw backs of the existing storage policies

The existing storage policies in a SAN (Storage Area Network) or NAS (Network Attached Storage) follows some already well established principles for loading the data with the data stores. Nearest geographically located server allocation, Round robin technique and Priority based allocation are some of the deployed principles [6] in assigning the server. But these algorithms are not taking the various influencing parameters into consideration.

Deployment of such algorithms may end up with poor utilization of storage resources as well as heavy taxing of some storage servers. Even though the accessibility is achieved with ease but it discards or give less preferences to the better performance issues which is a challenging quest when the system requires voluminous data for the user requests in the domain of warehousing and mining.

### 4.2 Need for Data Inventory Control

Ever growing nature of storage requirements [9] and the need to retrieve information without compromising the performance is a continual challenge and has been handled by adding more storage capacity and by adapting or introducing policies like users deleting or moving unnecessary information to different and less expensive storage tiers. This, 'most of the data, most of the time' requirement puts new demands on the storage infrastructure to follow better inventory decisions which allows the right data at right place but not in dumping everything at the same place without any principle.

For example, Fig. 1 represents the storage servers assignment for different sites of different users who are consuming storage as a service which is a common scenario seen in any of the cloud based storage service models.

Assume the principle of assigning users data in the data store may follow the principle of assignment of servers in famous storage service provider like Amazon S3 (Simple Storage Services). Amazon [9] made data stores available in eight regions namely US Standard, US West (Oregon), US West (Northern California), EU (Ireland), Asia Pacific (Singapore), Asia Pacific (Tokyo), South America (Sao Paulo), and Amazon Web Services GovCloud (US) to cater the storage requirements of their global users.

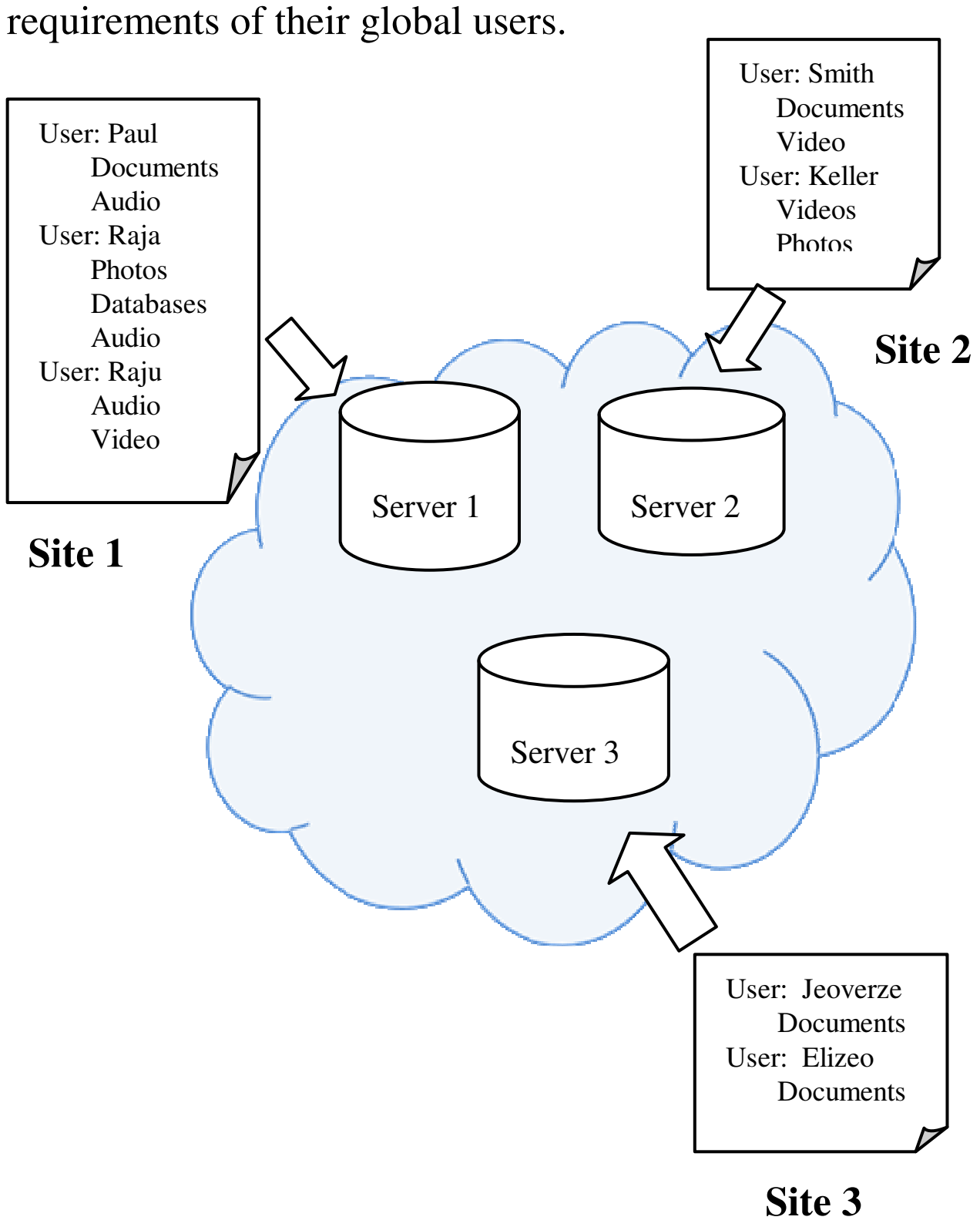

Fig. 1 Storage server assignments without data inventory principle

There are several factors to consider based on the users preferences on a specific application. The users may want to store their data in a Region that is near to them, their data centers in order to reduce data access latencies or that is remote from their other operations for geographic redundancy and disaster recovery purposes or enables them to address specific legal and regulatory requirements or they want to go for lower priced regions to save money [5]. The major frustrating inbounded nature of such a deployment of users' preferences will lead to poor performance even though it is the decision imposed by the user. Aiming for the data availability with the users preferences without compromising the performance will be a mirage if we do not rely ourselves with better inventory principles in terms of data management.

## 5. Principles of data inventory control

As the items or components or raw materials are consumed by the consumers in product inventory system, data is the only consumable commodity in the context of data inventory system. Users are ubiquitous in nature in almost all massive storage required applications which are accessed through the Internet and provision of improved performance will surely improve the satisfactory level of the users.

### 5.1 Data Inventory Policy

As the product inventory control focuses on how much stock of items should be maintained at hand as well as when it should be replenished, data inventory management should answer the following questions:

1. Optimum data that a data store can have for its storage requests from the users.
2. How much and what data has to be distributed or moved among different data stores based on the change in the storage requests.

### 5.2 What data should be kept in a data store?

The answer to the question of keeping the optimum data in a data store is determined by minimising the various cost parameters in the data management perspectives. There are various cost factors which influences the product inventory management. Similarly we are proposing the following as some of the costs factors to be considered if we develop data inventory cost models. These cost parameters can be equated to the inventory economic parameters so that the inventory models available in inventory management can be restructured with alterations fitting to the context of data inventory system.

The following are the some of the basic costs involved in the inventory management [5].

- Purchase costs - It is the price per unit of the item.
- Ordering costs - Costs involved in preparing the purchase order and invoices, stationery etc.
- Carrying costs - Costs of maintaining the inventory.
- Shortage costs - Cost incurred when there is no stock in hand.

As the product inventory control is having different costs, data inventory is also having many major impacting parameters which can be equated to the economic parameters of inventory management. The following are the most inevitable parameters in the domain of storage centric networked architectures:

- Number of Users: The existence of number of users across the network leads to the major decision of assigning the data stores appropriately. It is obviously having impact over the other impacting parameters like size of data, frequency of access, queue size etc.

- Frequency of Access: It refers the number of storage requests from the users as well as the data which is most frequently accessed.
- Size of the data: It refers to the data measured in terms of its volume. When it is classified based on some other parameters the resultant should lead to less fragmentation.
- Type of the data: Certain types of data need to be considered as the whole but not desirable to undergo distribution.
- Queue size: The number of users, their requests and frequency of requests decides the queue size. The performance is inversely proportional to this queue size.
- Nature of the requests: The primitive storage request Write will be costlier comparing with the request Read.
- Capacity of the data store: A group of heterogeneous data stores will be deployed with different hardware and software. Certain combination of the hardware and software profile will speed up the performance and vice versa.
- Geographical Location: Assignment of nearest data store will result with omission of network limitations.
- Fault tolerance: Backup and replication [3] policies are inversely proportional to the retrieval rate.
- User Control over the data: Setting up the preferences in selecting the data store or switching between the networks should be considered.
- Participation policy of the node: Due to reasons like security or predicted network interruptions, a node may isolate by itself.
- Network Cost &Traffic rate: Based on the band width, access time, queue size and the protocol which is deployed in the network, it will ease the performance.

For example, the carrying cost in inventory management is equivalent to the network cost and the traffic rate if these parameters are represented with proper metrics.

### 5.3 When should the data be moved among data stores?

The reorder level of an inventory is the point at which stock on a particular item has diminished to a point where it needs to be replenished. The reorder level of stock is often set at a figure higher than zero to take this time period into account. Therefore, the reorder level is set so that the stock level will reach at or around zero about the time the next shipment of stock is anticipated to arrive.

In the context of data inventory management, reorder point will be a conceptual phenomenon of sensing the data items needed for a data store corresponding to its users requests. Those data items which are sensed will be moving among the data stores and will be placed appropriately in the data tier architecture proposed in the section 5.4.

### 5.4 Data Management in storage servers with Inventory Control

This section proposes a data inventory model which is derived from the basic principle behind product inventory system. Data inventory model which is proposed here takes the parameters of the product inventory models and equates those parameters fittingly for the data management principles. This proposed model aims to result with right data at right place at right time.
In this model, the data that will be placed in a data store is evaluated based on various influencing parameters. Flow diagram in Fig. 3 represents the implementation of data inventory control in which the data is assigned to any of the specific data store.

Fig. 4 illustrates the classification of data availability in a data store and the data placement in the different levels based on its various impacting parameters which are listed in Sec.5.2.
For example a critical data or transactional data might be placed in the higher order level and the data which are infrequently accessed might be placed at the lower order level. Keeping the high order level data at the own custody of a data store will improve the throughput. The lower order level data might be made available at the nearest data stores or distributing those data among the other data stores and the selection of such data stores for distribution [12] will be again based on the impacting parameters. Whenever there is a change in the requests for that data, it will dynamically make changes in its order of placement.

This reordering of data items within the data store or among the data stores can be equated to the principle behind Reorder levels of product inventory. Thus this principle answers the first question of inventory policy of where to keep a data item.
A well exerted data inventory control should sense the data items which might be required for each and every data store well in advance. Once a specific data is found to be needed for replacement to another data store, there is a need for data distribution or data movement among the data stores. Such requirements of replicated data [20] can be equated to the Reorder levels in the context of product inventory.

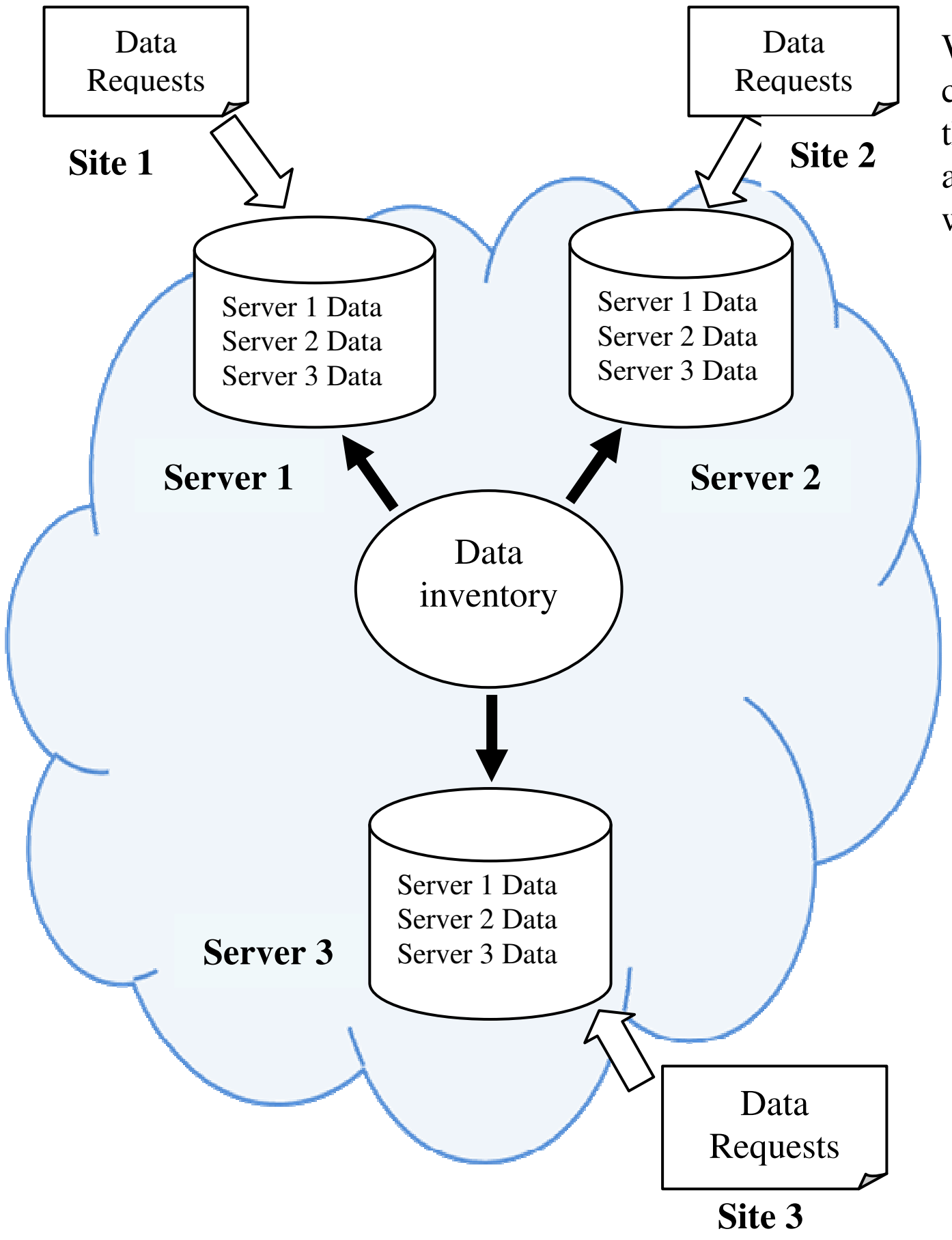

Fig. 2 Implementation of data inventory control resulting with the distribution of data among storage servers.

Thus there will be a dynamic evaluation in the data placements based on the users requests for data results with movement of data among the data stores or the data will change its resident of order of level. The data which is requested may be a replication or it may be the data shifted to some other order of level in the same data store.

Whatever be the principle behind the assignment of data stores for the users, there must be a proper envisioned storage management principle [8] which is similar to the management principles of product inventory management control. Devising and exercising such a data inventory principle would result with high performance and give a way to reach better satisfaction for the users. Fig. 2 represents the scenario of previously illustrated storage servers with the implementation of data inventory control.

For example, the carrying cost in inventory management is equivalent to the network cost and the traffic rate if these parameters are represented with proper metrics.

We can also classify the data inventory models into two categories based on the requirements from the users. If the users demand is static then those models are termed as 'Deterministic' else it is 'Probabilistic or Stochastic' where demand varies.

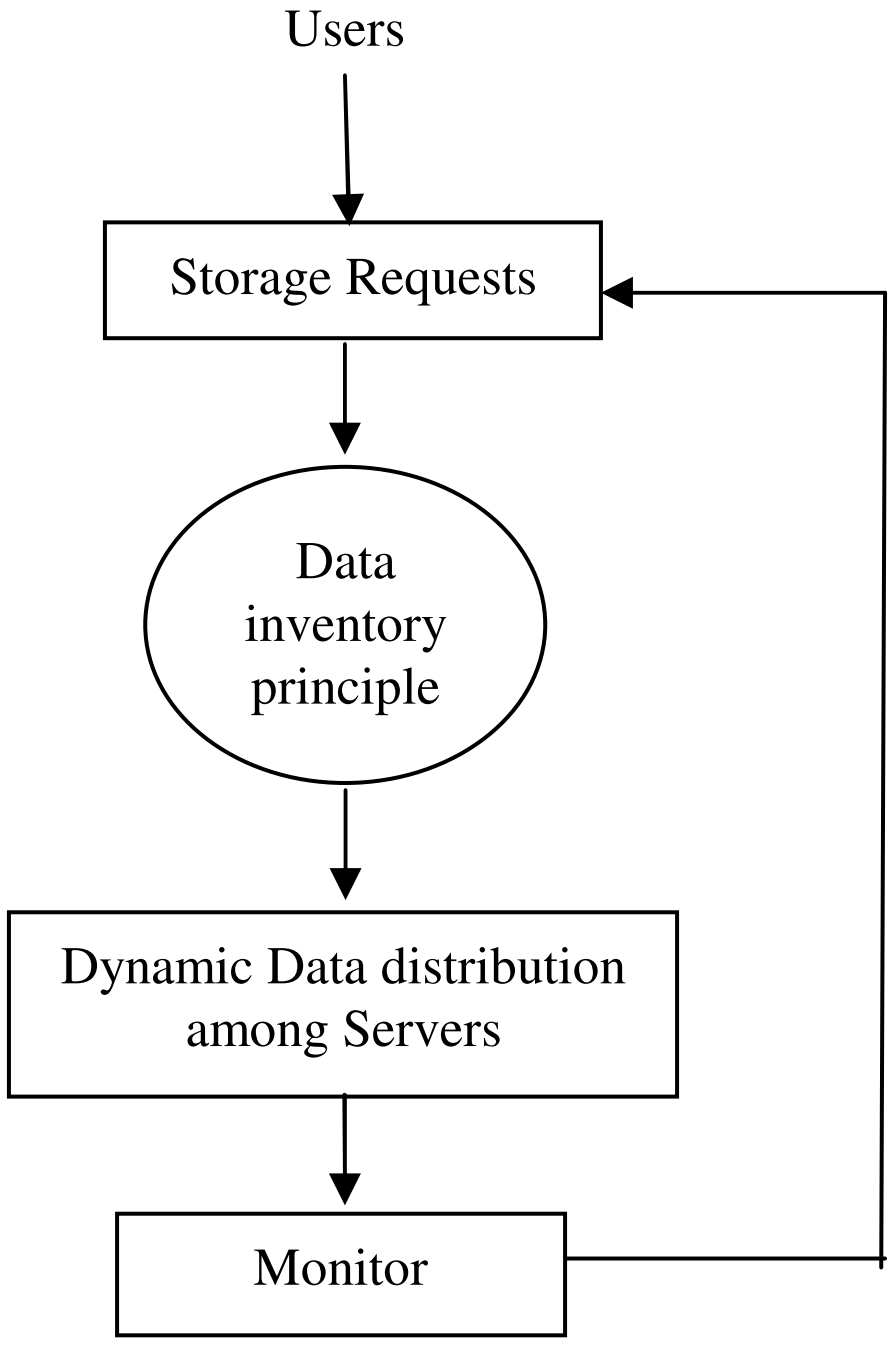

Fig. 3 Flow diagram represents exercise of Inventory control.

## 5.5 Does data inventory control tend to data volume load balancing?

Here the data inventory control is not trying to balance [14] the data in terms of volume which a particular storage server is possessing but in principle it tries to distribute the data [16] among themselves so that it leads to maintain the reposition of optimized data for the users' requests. But at situations of types of storage requests the data stores receive and the impacting parameters at that moment, it may lead to a balance among the data stores in terms of volume of data [10]. The data inventory control should consider and evaluate the various impacting parameters which can be equated to economic parameters of product inventory management.

### 5.5 Relevancy of data inventory control

- It maximises the performance of data retrieval much faster by placing appropriate data at appropriate place.
- Leads to better utilization of storage resources instead of investing on them.
- Scale up will be a costlier affair but it can be efficiently managed by scale out strategies. Data inventory control lead to scale out strategy when parameters are properly plotted.
- Big data management [12] at lower storage costs is possible to achieve.
- Scalability [3] and virtualization [17] of resources are the basic objectives of the booming business models like Cloud computing [13]. Data inventory control aims for virtualization of resources with cheaper cost by its principle but achieving it by scale up solutions will be costlier.

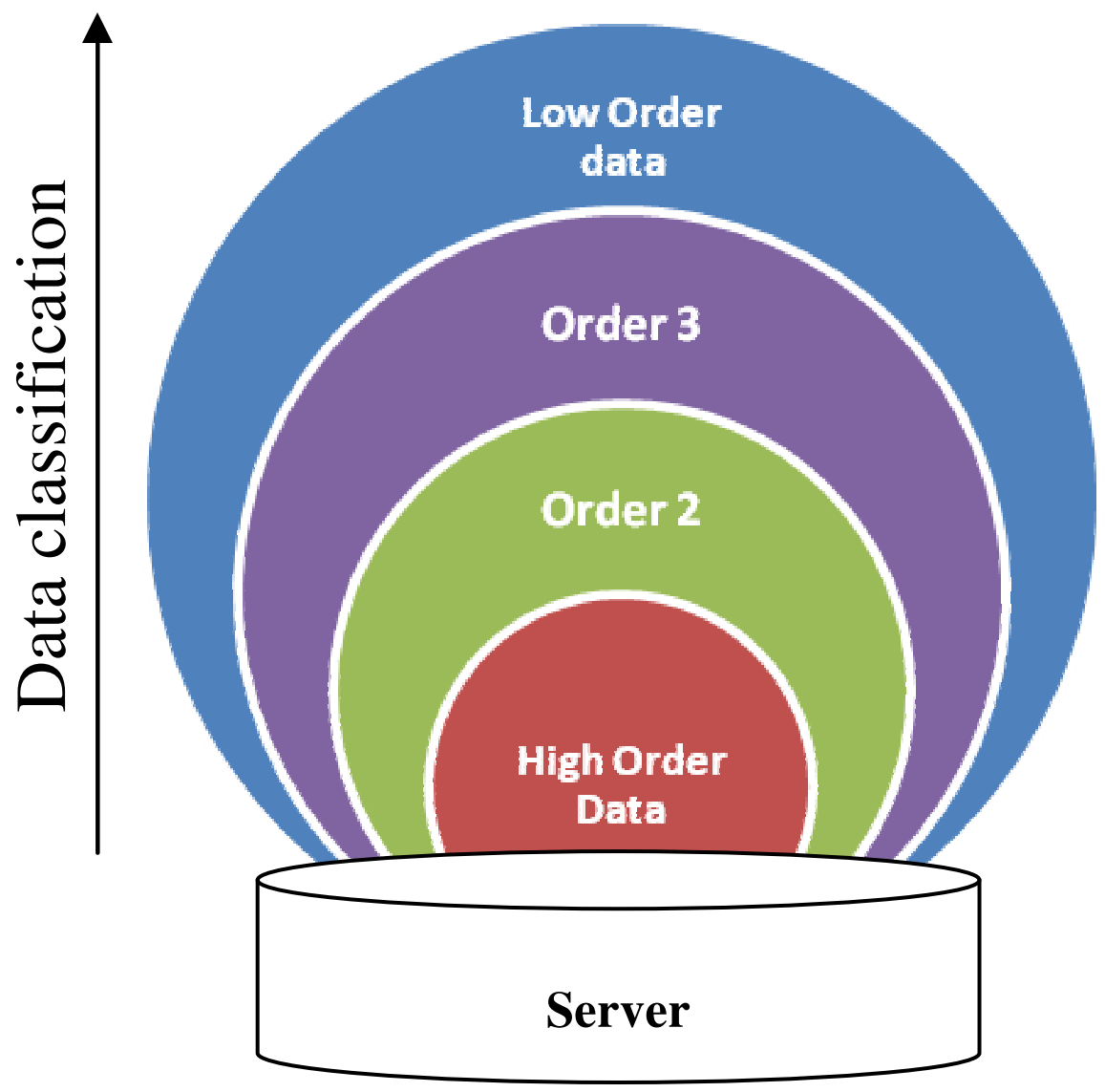

Fig. 4 Data classification and its placement in a data store

**R Arokia Paul Rajan** is working as Associate Professor in Pope John Paul II College of Education, Pondicherry as well as a Ph.D Scholar in the Department of Computer Science & Engineering at Pondicherry Engineering College, Pondicherry, India. He obtained his M.C.A (1999) from Bharathidasan University, India.

**F Sagayaraj Francis** is working as Associate Professor in the Department of Computer Science & Engineering at Pondicherry Engineering College, Pondicherry, India. He holds Ph.D in Data Management from Pondicherry University, Pondicherry, India. He has authored 10 journals and 8 conference publications. Some of his areas of interest include Database Management Systems, Data Mining and Knowledge Discovery, Data Structures and Algorithms, Knowledge and Intelligent Systems and Automata Theory and Applications.